\documentclass[reprint,amsmath,amssymb,aps,showkeys,floatfix]{revtex4-2}

\usepackage{graphicx}
\usepackage{dcolumn}
\usepackage{bm}
\usepackage{amsmath} 
\usepackage{braket}
\usepackage{ulem}
\usepackage{mathtools}
\usepackage[none]{hyphenat}
\usepackage{xcolor}%
\newcommand{\vb}[1]{\mathbf{#1}}
\newcommand{\p}[1]{\left( #1 \right)}
\newcommand{\br}[1]{\left[ #1 \right]}
\newcommand{\cbr}[1]{\left\{ #1 \right\}}
\newcommand{\abs}[1]{\left| #1 \right|}

\begin{document}

\preprint{APS/123-QED}

\title{Quantum Mechanical Description of Active Matter}

\author{Ghansham Chandel}
 \altaffiliation[Also at ]{Mechanical Department, University of Maryland at College Park.}
\author{Siddhartha Das}%
 \email{sidd@umd.edu}
\affiliation{%
Department of Mechanical Engineering,\\
 University of Maryland at College Park.
}%

\date{\today}

\begin{abstract}
The natural environments habitable to life are inseparable from dissipative and entropic forces, which mandate active matter to operate on constant fluxes of energy and momentum to survive and propagate through spacetime. In contrast to fundamental particles which are characterized by energy and momentum, we postulate that energy expenditure rate and momentum generation rate are the fundamental dynamical quantities driving this evolution of swimming active matter and accordingly construct the corresponding Lagrangian and Hamiltonian of motion of the active matter. Furthermore, living systems broadly employ periodic, non-repeating motion for both locomotion and various biological functions, with meaningful information stored in the periods of these functions. Consequently, analyzing these dynamics in the Fourier space lands us at the familiar tools of the operator based, quantum mechanics. This is accomplished by introducing a complex field to describe the evolving probability distribution of a single unit, as well as the population density of mutually non-interacting active particles under the constraint of scarce resources. The resulting single-unit wavefunction evolves analogously to the Dirac equation with minimal coupling, wherein gauge symmetry reflects the generality of the governing equations across arbitrary resource landscapes. Additionally, we present a Schrödinger-like equation where the effective potential term represents the environmentally mandated, non-motile base metabolic rate. Finally, we formalize the bidirectional coupling between active swimmers and the surrounding fluid via a multipole expansion of the active force with an emphasis on force monopoles and dipoles, where each pole corresponds to a distinct mode of active motility.

\end{abstract}

\maketitle
\section{Introduction}\label{sec1}

Following billions of years of evolution, the existence and activity of active creatures span extraordinary range of length scale, time spans, temperatures, and pressures among other environmental factors. Consequently, active matter community has identified a vast range of active behaviors such as run-and-tumble dynamics \cite{solon2015active}\cite{willems2025run}, polar flocks \cite{quillen2020boids}\cite{baconnier2025self}, dry \cite{ngo2014large} and wet \cite{thampi2016active} active nematics, to name a few \cite{marchetti2013hydrodynamics}\cite{Yeomans08122025}. The intersection of active matter and quantum mechanics has grown to bridge two main perspectives: classical active particles exhibiting quantum-like behavior, and quantum particles consuming energy actively. The examples of classical active particles modeled as quantum systems include polar particles showing spin-orbit coupling \cite{loewe2018flocking}, inertial models showing wave-like tunneling \cite{te2023microscopic}, heterogeneous nematics showing potential well dynamics \cite{houston2024spontaneous}, defects (pseudo-particles) with conserved topological charge undergoing pair creation/annihilation \cite{Yeomans08122025} and confined active particles showing Casimir-like fluctuation attractions \cite{ray2014casimir}. Conversely, the examples of quantum particles as active matter, spin possessing atoms show active kinetic extraction via spin exchange \cite{burgardt2026quantum}, and dissipative quantum systems show active noise \cite{antonov2026modeling} among many other active matter analogues. While these mappings reveal meaningful mathematical similarities, the literature that captures the ``drive'' of fundamental particles to minimize potential energy in a direct analogy with the living active matter which migrate for survival is elusive.

In this paper, we are specifically interested in active creatures that exist under scarce resources and lack freedom to spend their energies in any non-motion related processes. We call such creatures ``miserable'' as they must constantly plan and move in search of ``food'' in order to survive or risk halting by exhaustion indefinitely. We claim this constraint to be an idealized ground for analyzing the fundamental dynamics of active matter. Just as Newtonian physics relies on conserved mass, zero gravity, and the absence of dissipative forces to isolate the core dynamical laws, this strict energy-to-motion constraint strips away biological complexities to reveal the pure physical laws governing active migration.

The dynamics in our framework is characterized by the fact that the evolution of active creatures is governed not through spatial gradients in energy (force), but through spatial gradients in energy expenditure rate (resource availability). {In other words,} we consider {such} dynamics where the objects move in a fluid-filled spacetime, the energy-expenditure-rate generates the time evolution, and force (or momentum-addition-rate) generates spatial evolution of any given object existing under the framework of Active quantum-mechanics (AQM).

To a degree of success, barring the Wallstrom objection \cite{reddiger2023towards}, the Madelung formulation \cite{waegell2024toward} of quantum mechanics shows that, the Schrödinger  equation for non-relativistic particles maps onto a set of inviscid hydrodynamic Euler equations, known as the Madelung equations. Furthermore, as seen in analog black hole models, the high Reynolds-number (Re) acoustic waves can mimic causal waves in spacetime. We depart from these high-Re dynamics, extending the quantum analogy into the micro-fluidic, Stokesian regime. A fundamental divergence from the Madelung formulation is that our model governs active creatures within a fluid, as opposed to pure fluid matter alone. This exercise ultimately lands us at a framework that fundamentally contrasts the deterministic Newtonian mechanics. Rather than explicitly solving for non-linear, time-evolving trajectories, the kinematics and kinetics are expressed as a linear combination of Fourier eigenmodes—modes that remain stationary in time for a steady potential, thereby substituting the system's inherent non-linearities with a wave-like superposition.

\section{Quantizing Active Matter:}\label{sec2}
The motivation of this AQM framework is rooted in three fundamental observations regarding the dynamics of living systems. First, unlike fundamental particles that exist with constant energies (or mass) in free space, living organisms require constant energy expenditure to maintain a steady state. Second, in the absence of active propulsion (e.g., walking, swimming, or slithering), organisms inevitably halt to rest due to the presence of dissipative effects such as friction, viscous drag, and acoustic losses, thereby, increasing the entropy of the universe in accordance with the second law of thermodynamics. Finally, in various successful continuum and discrete flock models, active organisms swim with minimal Brownian diffusion. The dispersion in such models happen due to ballistic diffusion instead, where higher velocity particles disperse naturally from the slower particles. Throughout the paper, we shall refer to any creature(s) under the umbrella of AQM as active-particle(s) or AP(s). 

Corresponding to a given species of APs, we assign a complex number to every point in spacetime, which is the starting point of our theoretical framework. As complex numbers circle back to the same state after each $\Delta \theta=2\pi$ phase shift, the choice of a complex field is an appropriate as well as the simplest mathematical object to describe the abundant wave-like patterns (see Fig. \ref{nfig_1}a) existing among innumerable animal kingdoms at molecular, cellular, tissue, organ, and herd levels of scales. With kinesin motors propelling microtubules, microbes swimming through flagellating cilia and flagella, colon transporting digestive materials, animals undergoing slithering motion, etc., nature settled on a large number of repeating, non-reciprocating motions, broadly labeled as peristaltic motion. To tabulate these motions, we introduce a complex field $\psi$, such that:
\begin{equation}
    \label{wf_def}
    \psi(\vb{x},t)=R(\vb{x},t)\exp\p{i\theta(\vb{x,t})}.
\end{equation}
where, $R(\vb{x},t)$ is a real-valued scalar field characterizing the spatial distribution of the active matter (e.g., the number density of active cells or the probability distribution of populations). The corresponding conserved density is defined as $\rho_\psi:=R^2=  \psi^*\psi$, where $\psi^*$ is the complex conjugate of $\psi$. We assume the total amount of active matter to be conserved by eliminating the first-order source/sink dynamics such as cellular multiplication, procreation, or death. Under these conditions, if we define the flux-conserving derivative as $\text{d}_{t}^{\p{\mathbf{v}}}=\p{\partial_{t} + \nabla\cdot\mathbf{v}}$, with the total velocity of the AP as $\mathbf{v}$, we get: $\text{d}_t^{\p{\mathbf{v}}}\rho_\psi=0$ and  $\text{d}_t^{\p{\mathbf{v}}}R=0$. This purely real-valued conservation law precisely mirrors the species transport equation under fluid advection with zero diffusivity and sources. Neglecting diffusivity is physically justified: if $\rho_\psi$ represents a single AP, structural rigidity prevents indefinite Brownian dispersion. If $\rho_\psi$ represents a population, we restrict our scope to cohesive flocking behaviors that actively penalize density fluctuations, thereby suppressing macroscopic diffusion \cite{reynolds2022stochastic}, i.e., one flock is one unit.

Next, $\theta(\vb{x},t)$ in Eq.~(\ref{wf_def}) denotes a real-valued complex phase field. Complex phase encodes and unifies the relevant classical quantities in common units of energy consumption rate. Consequently, the complex wavefunction enables us to infer that any factor that modulates the metabolic rate of APs will directly govern the APs' physical dynamics, provided the APs exhibit a drive and means to migrate to nutrient-rich, or equivalently, less toxic environments. Corresponding to many metabolic rate influencing factors such as temperature, chemical species, radiation, pressure, etc. that can be easily arranged to be spatially varying, we have APs with adaptations such as thermotaxis \cite{garrity2010running}, chemotaxis \cite{wadhams2004making}, phototaxis \cite{wilde2017light}, rheotaxis \cite{nakane2023rheotaxis}, etc. informing and operating on gradients of said signals. We propose a simple model where the utility function of the creatures is to maximize the difference between available resources (power consumption sustained by the environment) and base-metabolic-rate (BMR, which is the motionless expenditure rate).

\subsection{Interpreting the Complex Phase Angle}\label{sub_CPA}
\begin{figure*}[htbp]
\centering
\includegraphics[width=0.98\textwidth]{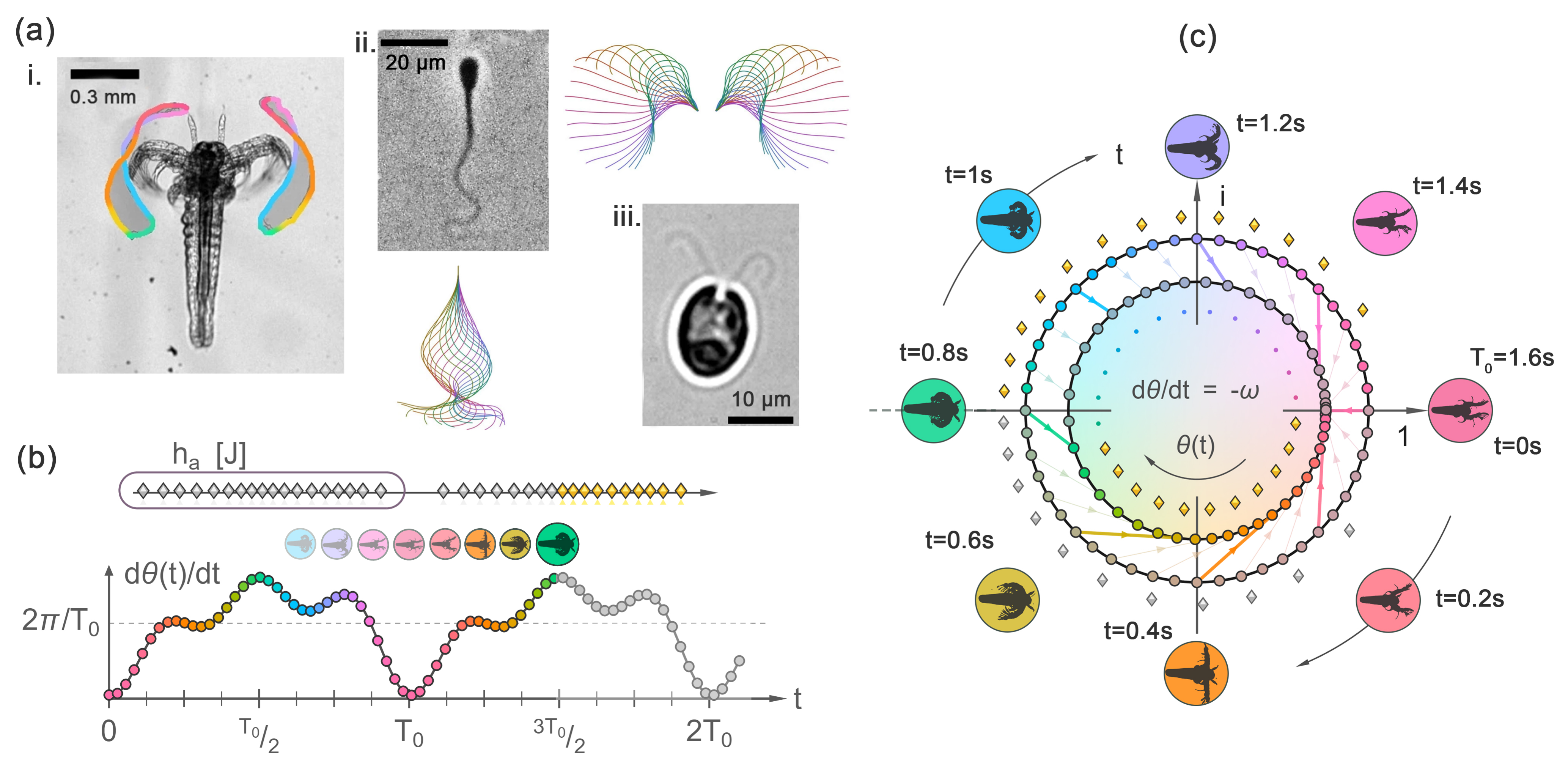}
\caption{Illustratively, this figure shows the concept of active creatures evolving through periodic, non-reciprocal motion. (a) Shows [i] \textit{Artemia}, Brine Shrimp (modified with {permission from} \cite{lara2026forces}) [ii] Sperm cell {permission from  \cite{walker2020computer}}, and [iii] Algae, \textit{Chlamydomonas reinhardtii} (modified with {permission from} \cite{moreau2024minimal}\cite{wan2014rhythmicity}). {In all the examples, the active creatures are} swimming in fluid in Stokesian regime and their respective trajectories {are shown} overlaid in the body center reference frame. (b) We illustrate the rate of evolution of the complex angle $\text{d}\theta/\text{d}t=-\mathcal{E}/\hbar_a$ as a function of time, with the fundamental period of motion as $T_0$ and the total energy spent in one period as $\hbar_a$. (c) Shows the mapping between constant time flow in the outer circle and a constant active-action $S_a$ in the inner circle. When energy-expenditure-rate $\mathcal{E}(t)=\hbar_a\omega(t)$ is slower, the complex phase $\theta$ is densely packed with the time snapshots, whereas when $\mathcal{E}$ is faster, the phase updates quickly. The diamonds illustratively represent a constant fraction of the active action $S_a$ which are equally spaced in the complex phase but are densely packed at times when the AP consumes energy faster.
}
\label{nfig_1}
\end{figure*}

We encourage one to assume a plane-wave solution, i.e., $\psi=\exp i\p{\vb{k}\cdot\vb{x}-\omega t}$, to harness intuition behind various definitions laid in this section. Physically this means that we have perfect knowledge about the momentum of our AP and there is no constraint on its position. Here, the kinematic quantities, $\omega$ and $\vb{k}$ are the angular velocity and wavenumber associated with the wavefunction. We shall try to answer what would be the evolution of the probability (or number-density) distribution, as well as, what more do we gain from evolving a square-root of the probability, and ultimately, give a quantum-mechanical description of active matter.

As illustrated in Fig. \ref{nfig_1}(b,c), with the proportionality constant $\hbar_a$ as the reduced Active-Planck constant, we propose the following mapping between the physical, possibly peristaltic and periodic motion of the APs and the action $S\ [\text{J-s}]$ equivalent of AQM, which is the activity $S_a=\hbar_a\theta(\vb{x},t) [\text{J}]$: the rate of change of $S_a$ is directly proportional to the magnitude of energy-expenditure-rate $\mathcal{E}(\vb{x},t)=\hbar_a\omega$, i.e.,
\begin{equation}
    \label{time_gen}
    \begin{split}
    \partial_t\theta(\vb{x},t)= -\omega(\vb{x},t) \\
    \partial_tS_a(\vb{x},t)=-\mathcal{E}(\vb{x},t).
    \end{split}
\end{equation}

\begin{figure}[htbp]
\centering
\includegraphics[width=0.48\textwidth]{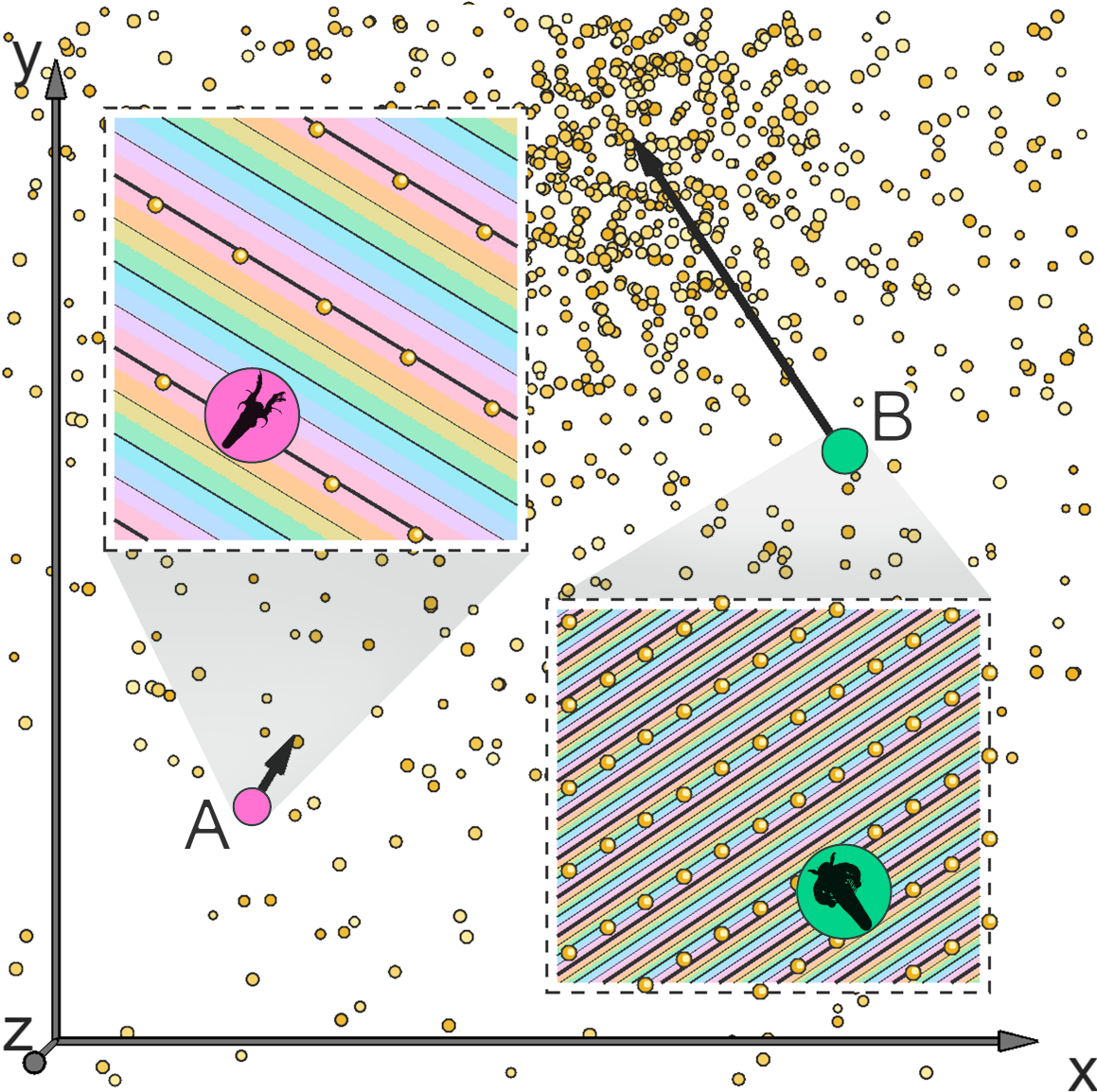}
\caption{ 
Illustratively, we show how the spatial distribution of the resources mandates the speed of the APs. Each scattered yellow circles represent a quanta of energy $\hbar_a$ as a discrete chunk. In our theory, this concentration is coarse-grained to a continuum which is proportional to $\phi(\vb{x},t)$. Active particle A is located in a relatively scarce environment in comparison to Active Particle B. Consequently, A must swim with lesser propulsion force $\vb{P}$ to conserve its energy to cover a greater distance as compared to B. Corresponding to each AP, we show the associated complex phase distribution locally with the inset detail-view sub-panels. Each complete spectrum represents consumption of one quanta of energy and one boost across space and time, which is inversely proportional to the wavelength and angular frequency of motion, and hence spatial and temporal gradient of active action $S_a$, respectively.
}
\label{nfig_2}
\end{figure}

In Eq.~(\ref{time_gen}), we see the kinematic and dynamic version of the same time generator equation. Assume the fundamental period of motion as $T_0$. Accordingly, integrating Eq.~(\ref{time_gen}) from $t=\br{t_0,t_0+T_0}$ such that $\theta$ goes from $\theta_0 \rightarrow\theta_0-2\pi$, we get $\int_{t_0}^{t_0+T_0}\mathcal{E} \text{d}t=2\pi \hbar_a=h_a$. Realize that for any periodic motion with period $T_0$, the integral of $\mathcal{E}$ is constant over that period. This gives us the definition of the of the Active Planck constant ($h_a=2\pi \hbar_a$) as the total energy expended in one step (period). Equivalently, $\theta(\vb{x},t)/2\pi$ is the number  of available quanta (packets of energy $h_a$ [J]) that the AP consumes with time if it was placed at $\p{\vb{x},t}$.

It is worth pondering the meaning of the fundamental period $T_0$. Truly returning to the same state after a fixed period includes losing all energy one gathered, i.e., the rate of gathering and rate of consuming resources must balance one another. We shall call such creatures as the ``miserable-'' APs, because expending more energy than available at each step (consuming one quanta) risks a halt to rest indefinitely, whereas spending less energy to reach to next quanta by moving slow implies an increase in stored energy or an increased BMR, which the creature must minimize according to our postulates. Next, the generator of translation, the momentum-production-rate $\vb{P}$ (or propulsion force, and not the momentum $\vb{p}$)  can thus be defined in a straightforward manner as:

\begin{equation}
    \label{pos_gen}
    \nabla\theta(\vb{x},t)=\vb{k}(\vb{x},t)=\vb{P}/\hbar_a.
\end{equation}
where ${\vb{k}(\vb{x},t)}$ is the local wavenumber of APs, with the appropriate illustration demonstrated in Fig. \ref{nfig_2}. From fluid mechanics, the force $\vb{P}$ acting on a body moving at its terminal velocity $\vb{v}_a$ is dictated by its drag coefficient $c_d$ as $\vb{P}=m_a\text{d}\vb{X}/\text{d}t=c_d{\vb{v}_a}$ instantaneously at low Re regime. Notice, the inertial mass has negligible effect on the dynamics and the effective mass $m_a$ of the AP is exactly the drag coefficient, i.e., $m_a=c_d$. From energy conservation, $\vb{P}\cdot\Delta \vb{x}=\mathcal{E}T_0=h_a$. This demonstrates the tradeoff for the miserable-AP when choosing to move, the motion can occur through higher propulsion and shorter distance, or low energy-expenditure-rate, longer time and a greater distance. Clearly, Eqs. (\ref{time_gen},\ref{pos_gen}) show us that energy-expenditure-rate and momentum-production-rate have seized to represent power and force and instead represent ``energy'' and ``momentum''. Next, we split the complex phase into two distinct fields.

\begin{equation}
    \label{phi_split}
    \theta(\vb{x},t)=\frac{q}{\hbar_a}\phi(\vb{x},t) +\theta_0\p{t} .
\end{equation}

Here, $\theta_0$ is the field that, does not change spatially but represents how an AP responds to changes in internal temperature, pressure, concentration of electrolytes, stress levels, or any other factors that affect the energy-expenditure-rate. Contrastingly, $\phi(\vb{x},t)$ is the energy available per unit charge which may be utilized by the AP if it were placed at any point in the domain. Interestingly, $q$ that acts as the electrostatic charge is the constant that the AP assigns to various resource-fields to scale the signal strengths. For example, if the fluid bath is laden with spatiotemporally varying concentration of glucose and adenosine-triphosphate (ATP), the AP must assign weights proportional to the ability of each resource to provide energy to it, i.e., scale their calorific values. A negative $q$, on the other hand, implies that the field represents toxins which the AP must avoid. 

In such frameworks, angular speed and wave-vectors provide the kinematic description of the active creatures, whereas, energy-expenditure-rate and propulsion force provide the kinetic/dynamic description of the same dynamics. The dynamical equation of free particle, which has fixed energy-expenditure and propulsion-force from our definitions laid in Eqs.~(\ref{time_gen},\ref{pos_gen}) become $\psi=\exp \br{i\p{\vb{P}\cdot\vb{x}-\mathcal{E}t}/\hbar_a}$. Similar to traditional quantum-mechanics, the operators that return the expectation value of ``energy'' $\braket{\mathcal{E}}=\bra{\psi}\hat{\mathcal{E}}\ket{\psi}$ and ``momentum'' $\braket{\vb{P}}=\bra{\psi}\hat{\vb{P}}\ket{\psi}$ are $\hat{\mathcal{E}}=i\hbar_a\partial_t$ and $\hat{\vb{P}}=-i\hbar_a\nabla$, respectively.

\subsection{The First Order in Space-Time, Fluid Mechanical Coupling}\label{sub_PD}

Up till the previous section, the wavefunction could very well be defined for any continuum that supports the notion of active creatures with velocities. However, as such, only the effect of the environment on the AP is captured. We improve on this by defining the total velocity of the APs in the external reference frame as $\mathbf{v}=\mathbf{v}_a+\mathbf{u}$, where $\mathbf{v}_a(\vb{x},t)$ is the active velocity of the APs relative to the fluid flowing with advection velocity $\mathbf{u}$. We must ask: if $\mathbf{v}_a$ is the velocity of the AP, how can it be a continuous field? This is possible because $\mathbf{v}_a$ represents the velocity that the AP could take if it was located at a point $(\vb{x},t)$, as dictated by the energy supplied by the resource fields, thereby, allowing for a probabilistic description of active-matter physics. Along a trajectory, recall that $\text{d}_{t}^{\p{\vb{v}}} R=0$ and thus the change in $\psi$ in ground-frame can be written as (please see the Supporting Information or SI for a detailed derivation):

\begin{equation}
    \text{d}_{t}^{\p{\vb{v}}}\psi= i\p{    \frac{q}{\hbar_a}\text{d}_{t}^{\p{\vb{v}}}\phi      +   \text{d}_{t}^{\p{0}}\theta_0  }\psi 
    .\label{phi_rate_1}
\end{equation}

Remember, when fields, e.g., $\phi, \theta_0$ or floating species $\Phi$ (here we introduce $\Phi$ as an arbitrary, different resource field) evolve in the arbitrary frame, the velocity that dictates the change in the field values is not the absolute velocity (velocity relative to ground-frame), but the velocity of the AP, relative to the field $\text{d}_t^{\p{\Delta\vb{v}}}\Phi$ . The non-motion related energy-expenditure field ($\theta_0(t)$) is associated with the AP itself and thus, its relative velocity is always zero. Accordingly, applying chain rule over the trajectory on $\phi$ gives us  (please see the SI for a detailed derivation):

\begin{equation}
    \p{\partial_{t} + \nabla\cdot\vb{v}} \psi= i\p{-\omega_0 - \omega + \vb{v}\cdot\vb{k}}\psi
    \label{q1}
\end{equation}

With the definitions suggested by the Eqs. (\ref{time_gen},\ref{pos_gen}),  the frequency of the wavefunction and the direction of motion of the APs is given by $\omega, \omega_0$ and $\vb{k}$. Notice if $\theta(\vb{x},t)$ is purely real in Eqs. (\ref{phi_rate_1},\ref{q1}), we can infer that the rate-of-change of $\psi$ is purely imaginary, i.e., the evolution is unitary and $\rho_\psi$ is conserved at all times. Next, we define mass density as $\rho$ and background velocity as $\bf{u}$ and describe their transport through space and time through the Navier-Stokes momentum balance equation as:
\begin{equation}
\partial_t \p{\rho\vb{u}} + \nabla\cdot\p{\rho\vb{uu}}
=-\nabla p + \mu_s\nabla^2\vb{u}+\mu_b\nabla\p{\nabla\cdot\vb{u}}+\vb{f}_a
.\label{NS2}
\end{equation}

In Eqs. (\ref{NS2}), $\mu_s$ is the shear viscosity, $\mu_b$ is the bulk viscosity, $p$ is the pressure field, and $\vb{f}_a$ is a general force per unit volume. We retain the nonlinear convection terms in Eq.~(\ref{NS2}) not because they are relevant to microscale dynamics, but because they elucidate the underlying transport structure governed by the operator $\text{d}_t^{\p{\vb{u}}}$. Additionally, the pressure relates to the mass-density at constant entropy via ${\partial p}/{\partial \rho}|_S=c^2$, where $c$ is the speed of sound. This pressure - density relationship becomes important when investigating dynamics over very short time scales and large distances, which is broadly, the domain of acoustics. Information generated at a specific point does not propagate instantaneously; instead, it travels within a `sound cone' in spacetime. Consequently, sound could also represent the field $\phi$ for complex APs which ``hunt'' through audio signals if it's capable of storing energy over long distances. Such APs, however, break our miserable-AP postulate which requires the balance of energy consumption and expenditure at each step. While many facets of such APs are expected to remain identical with the miserable-APs, the creatures that can store energy indefinitely are beyond the scope of this paper.

Next, we consider $\vb{f}_a$ in Eq.~(\ref{NS2}), and assume $l_a$ to be the effective length of an individual AP. Although the active force operates as a continuous field distributed over the fluid, analyzing the long-range hydrodynamic impact of a discrete AP requires localizing its overall influence around a specific reference point: the particle's center of action, defined as $\vb{X}$. If the AP's effective length is much smaller than the relevant characteristic length scales (i.e., $l_a \ll v_a T_0$), we can rigorously approximate its far-field contribution by carrying a multipole expansion of the active force as (see the SI for a detailed derivation):
\begin{equation}
\label{point_MP}
\begin{split}
    (f^i)_a(\vb{x}) &= F^i \delta(\vb{x} - \vb{X}) - D^{ij} \partial_j \delta(\vb{x} - \vb{X}) \\
    &\quad + \frac{1}{2} Q^{ijk} \partial_j \partial_k \delta(\vb{x} - \vb{X}) - \dots
\end{split}
\end{equation}

In Eq.~(\ref{point_MP}), the vectors and tensors are expressed in index notation and summation over repeated indices is implied. In Eq.~(\ref{point_MP}), $\vb{F}=F^i \vb{e}_i$ is the prescribed coordinate frame with unit directions $\vb{e}_i$ and represents any external force acting on the AP, commonly known as the force monopole moment. This term is typically zero for force-free active microswimmers suspended in a fluid. However, if an AP is electromagnetically charged, has non-negligible buoyancy or gravitational mass, or pushes or pulls on objects that are stationary relative to an outside observer (such as walls, ladders, webs, or microtubules), $\vb{F}=c_d\vb{v}_a$ represents this external force. Similarly, in Eq.~(\ref{point_MP}), $\vb{D}=D^{ij}\vb{e}_i\vb{e}_j=c_d\p{\vb{v}_a\cdot \vb{l}_a} \vb{e}_a\vb{e}_a$ \cite{Lauga_2020}\cite{ganguly2025hydrodynamics}\cite{zottl2023modeling}  is the force dipole term, where $\vb{e}_a=\vb{v}_a/\abs{\vb{v}_a}$ is the unit vector in the direction of the active-velocity (or the wavevector) of the AP, as defined by Eq.~(\ref{pos_gen}). Here, the effective length $l_a$ represents the distance between the center of propulsion and the center of drag. The dot product of this dipole length, which is a vector directed from the center of propulsion to the center of drag and the active velocity, $\p{\vb{v}_a\cdot \vb{l}_a}$, is positive for pullers and negative for pushers. If we define the distance from the AP's center to its center of propulsion as $l_1$, and to its center of drag as $l_2$ (note that $l_1 + l_2=l_a$ for APs symmetric about the axis of motion), the quadrupole moment is in Eq.~(\ref{point_MP}) given by $\vb{Q}=Q^{ijk}\vb{e}_i\vb{e}_j\vb{e}_k=c_d v_a \p{l_2^2-l_1^2}\vb{e}_a\vb{e}_a\vb{e}_a$ \cite{Lauga_2020}\cite{ganguly2025hydrodynamics}\cite{zottl2023modeling} .

Notice that each term in Eq.~(\ref{point_MP}) is proportional to the active speed $\vb{v}_a$, which from Eqs. (\ref{pos_gen},\ref{phi_split}) becomes proportional to the charge $q$ and the number of quanta $\phi/\hbar_a$.
To generalize Eq.~(\ref{point_MP}) to a system of multiple non-interacting particles, or to model a single AP with discrete zones of force production (e.g., multiple fins or fine ciliary hairs distributed over its body), one can sum over all APs (or all propulsion points) as $\vb{f}_a=\sum \vb{f}_a^{(n)}$. This summation commutes with the partial derivatives subsequently coarse-graining the Dirac delta functions into a continuous AP density field, $\rho_\psi$, yields:

\begin{equation}
    \label{dist_MP}
    (f_i)_a(\vb{x}) = q\tilde{F}^i\rho_\psi - q\tilde{D}^{ij} \partial_j \rho_\psi + \frac{q}{2} \tilde{Q}^{ijk} \partial_j \partial_k \rho_\psi - \dots
\end{equation}
Here, $\cbr{\tilde{\vb{F}}, \tilde{\vb{D}}, \tilde{\vb{Q}}}=\cbr{\vb{F},\vb{D},\vb{Q}}/q$ are defined as such to be a constant over any and all resource field by scaling with the affinity of APs, namely, the field specific charge constant $q$. Additionally, Eq.~(\ref{dist_MP}) provides a framework that not only describes the transition between single and multiple AP systems, but also resolves the near-field flow around complex APs. Because $\rho_\psi$ is an advected field, $\vb{f}_a$ inherently advects not just with the background flow ($\vb{u}$), but with the active swimmers themselves (i.e., the total velocity of APs, $\vb{v}=\vb{u}+\vb{v}_a$). The continuity equation of Navier-Stokes remains untouched as:

\begin{equation}
\nabla\cdot{\vb{u}}= 0
,\label{NS1}
\end{equation} 

If the non-motion related energy-expenditure-rate $\hbar_a\omega_0$ is directed towards cellular multiplication (such as muscle, healthy cells or tumor growth), or reproduction, or simply lungs compression and expansion while breathing (puffers), the corresponding change in the velocity field may be added as a source monopole, i.e., the first polar moment of the $\dot{\rho}_\psi$ distribution. Higher order poles of active-density-rate distribution are comparable to the poles of force distribution of one order less, i.e., source dipole is comparable to the force-monopole, source-quadrupole is comparable to the force-dipole, etc., and the general trend is captured by the Lighthill's analogy \cite{lighthill1952sound}\cite{lighthill1954sound}. Since we model active-swimmers dynamically, we do not model the kinematic (volume) source/sink ($\propto\dot{\rho}_\psi$), which is enforced to be zero in this paper.

Yet again, multi-pole expansion reinforces the concept of ``charge'' associated with APs. Furthermore, Eq.~(\ref{q1}) is symmetric under local $U(1)$ phase transformation -- one must ask, what does this gauge symmetry represent? By defining the four-velocity as $v^\mu=\cbr{ct,\vb{v}}$ and the four-gradient as $\partial_\mu=\cbr{\partial_t/c,\nabla}$, Eq.~(\ref{phi_rate_1}) can be rearranged to obtain:

\begin{equation}
    \p{\, \partial_\mu v^\mu+ i \frac{q}{\hbar_a}v^\mu \partial_\mu\phi + i\omega_0}\psi=0. 
    \label{gauge}
\end{equation}

Eq.~(\ref{gauge}) in conjunction to it's coupling with Eq.~(\ref{NS2}) reveals a clear resemblance between the evolution of our wave function and the Dirac equation. In the far field of a single AP, the velocity field is divergence-free; therefore, the four-velocity commutes with the four-gradient. Such framework perfectly parallels the gauge-theoretic description of particles interacting with an electromagnetic field (EMF) \cite{griffiths2008introduction}. Meaningfully, we can define the covariant derivative $D_\mu=\partial_\mu+i {q}A_\mu$ and a motive field $A_\mu=\partial_\mu\phi/{\hbar_a}$. Similar to the gauge theoretic description of the EMF, $A_\mu$ is the addition needed to enforce local $U\p{1}$ phase symmetry to the active wavefunction. Furthermore, whenever a new resource field $\Phi$ of affinity $q_\Phi$  is added to the system, the motive field must transform as $A_\mu \rightarrow A_\mu+ q_\Phi\partial_\mu \Phi/\hbar_a$ to guide the AP to move towards the newly added resources, all while conserving the power-expenditure and force-generation in the system. Additionally, $\hbar_a\omega_0$ is the BMR, which is the energy-expenditure unrelated to the dynamical motion. 
\begin{figure*}[htbp]
\includegraphics[width=0.98\textwidth]{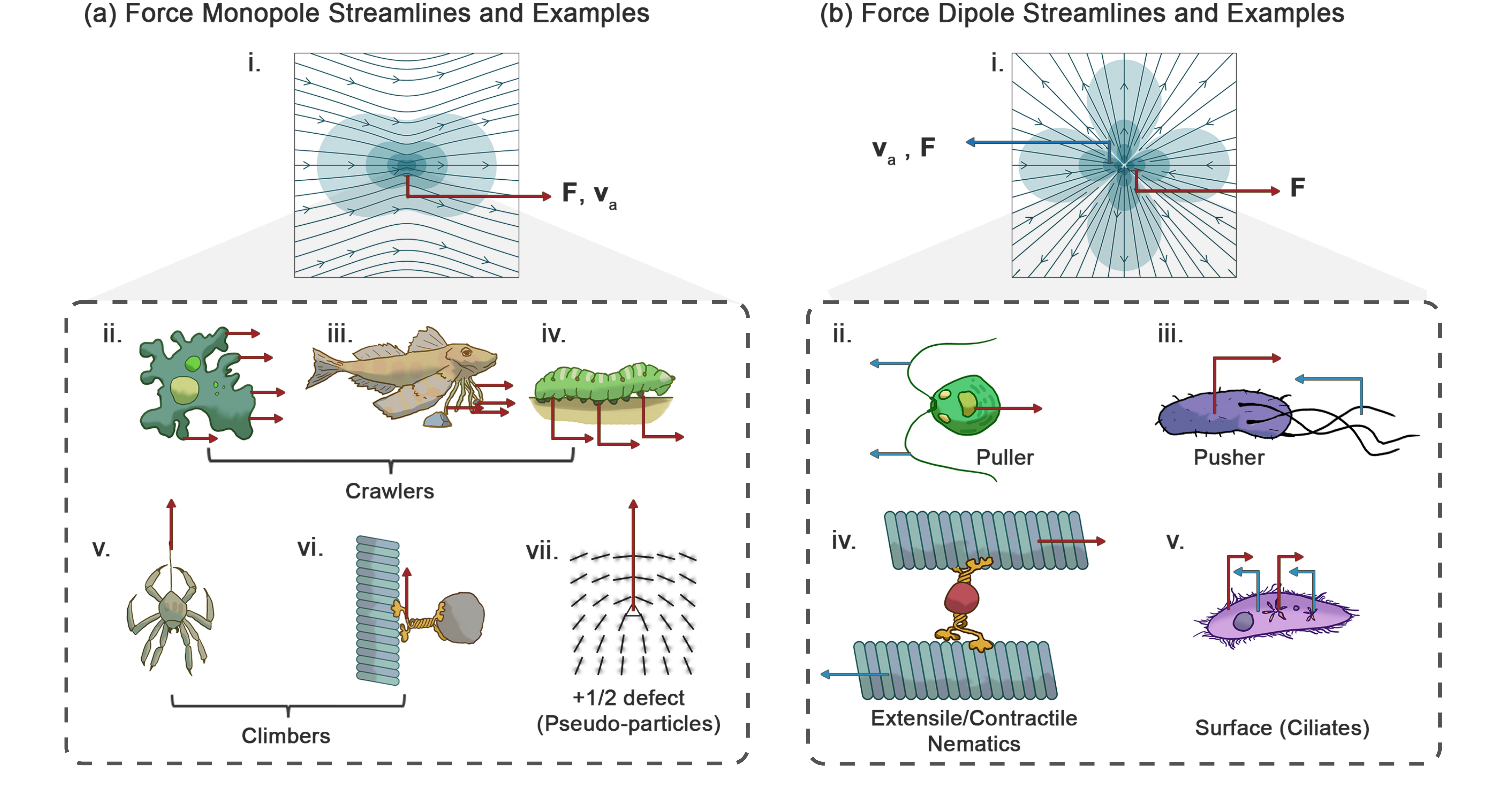}
\caption{
(a-[i]) We show streamlines for flow around a force monopole in Stokesian regime in an infinite domain: $\mathbf{u}_{\text{monopole}}(\mathbf{x}) = \mathbb{G}(\mathbf{x}) \cdot \mathbf{F}$, where $\mathbb{G}(\mathbf{x}) = \frac{1}{8\pi\mu_s x} \left( \mathbf{I} + \frac{\mathbf{x}\otimes\mathbf{x}}{x^2} \right)$ is the Oseen tensor, or the effective Green's function for Stokes flows. (a-[ii-vi]) We illustrate various biological entities that rely on an external force to propel forward, namely [ii] amoeba, [iii] Northern Searobin or \textit{Prionotus carolius} (a bottom-dwelling marine fish) [iv] generic centipede [v] generic spider and [vi] Kinesin motor transporting cargo by climbing on microtubeles. [vii] Shows +1/2 defect in active extensile nematics propelled forward with a force monopole without any external contact. (b-[i]) Shows streamlines for a symmetric force dipole: $\mathbf{u}_{\text{force-dipole}}(\mathbf{x}) = - (\mathbf{l}_a \cdot \nabla) \mathbb{G}(\mathbf{x}) \cdot \mathbf{F}$. (b-[ii-v]) Shows typical examples for force-free swimmers and force dipoles, namely [ii] puller Algae, \textit{Chlamydomonas reinhardtii}, [iii] pusher \textit{E. coli} [v] Kinesin cluster propelling a pair of microtubules, [iv] surface cilia hair driven ciliate, \textit{Paramecium}.
}
\label{nfig_3}
\end{figure*}

\subsection{The Active Hamiltonian}\label{sub_TAH}

The active action, or the activity $S_a$, as seen from Eqs. (\ref{time_gen},\ref{pos_gen}) is built from two quantities, which are namely, the propulsion-force $\vb{P}$ and the total energy-expenditure-rate $\mathcal{E}$ and their conjugate variables, namely, position $\vb{x}$ and time $t$. 

\begin{equation}
    \label{def_activity}
    S_a=\int^t_0\p{\vb{P}\cdot \dot{X}-\mathcal{E}}\text{d}t.
\end{equation}

Notice that in Eq.~(\ref{def_activity}), for a ``free'' AP, the resource landscape is uniform and $\dot{\vb{X}}\neq\dot{\vb{X}}(t)=\text{const.}$, and thus, the plane wave solution, $\psi=\exp \br{i\p{\vb{P}\cdot\vb{x}-\mathcal{E}t}/\hbar_a}$, is straightforwardly restored from the definition $\theta=S_a/\hbar_a$. From classical mechanics, we know that the integrand of the action (activity) is the Lagrangian, the Legendre factor is $\vb{P}\cdot\dot{\vb{X}}$ and thus, the constant of the motion, the Hamiltonian is $\mathcal{E}$. Physically, this means that when a miserable-AP evolves, its net energy-expenditure-rate remains constant. If the environment demands higher BMR or has scarce resources, the AP must compensate by reducing the propulsion force and swim slower to cover greater distances until it reaches the next food source. Let us call the difference between the BMR and the expenditure-rate permissible by the environment as the activity-potential $\mathcal{U}(\vb{x},t)$. Using the commutativity of time and space derivatives in Eqs. (\ref{time_gen},\ref{pos_gen}), we get the first Hamilton's equations:

\begin{equation}
    \label{pow_force}
        -\nabla \mathcal{E}= \frac{\partial\vb{P}}{\partial t}
        .
\end{equation}

Clearly, the ``motivation-force'' $\mathcal{F}=-\nabla \mathcal{E}$ is the gradient in the energy-expenditure-rate. This motivation-force from Eq.~(\ref{pow_force}) equals $\mathcal{F}=c_d\vb{a}_a$, i.e., effective mass times effective acceleration for constant a drag coefficient. Realize that the total-energy-expenditure-rate (the active Hamiltonian) is a function of position and momentum, i.e., $\mathcal{E}=\mathcal{K}(\vb{P})+\mathcal{U}(\vb{x},t)$. Here, the power used by the AP to move is represented by the kinetic-power $\mathcal{K}$ and the power not (directly) useful for motion is represented by $\mathcal{U}$. 
To establish the exact form of the kinetic-power, we invoke Hamilton's second equation. By definition, the canonical velocity of the AP is given by $\vb{v}_a = \partial \mathcal{E} / \partial \vb{P}$. Given our low-Reynolds number, $\vb{P} = c_d \vb{v}_a$; accordingly, on integrating with respect to the momentum-production-rate straightforwardly yields the kinetic-power term $\mathcal{K}(\vb{P}) = \vb{P}^2/2c_d+\mathcal{K}_0$. The constant of integration, being independent of momentum, represents the power consumed at rest: the base-metabolic-rate (BMR) $\mathcal{U}(\vb{x},t)$. 

Physically, if an AP is at rest initially, $\vb{P}\p{\vb{x}_0,t_0}=0$, the entire energy consumption goes solely to the base-metabolic-rate, i.e., $\mathcal{E}(\vb{x}_0,t_0) = \mathcal{U}(\vb{x}_0,t_0)$ and $\mathcal{K}_0=0$. There are two motivating factors that could force the AP to move: first, it could move to a less hostile environment where the new BMR is lower than the starting position; or second, it moves to a more fertile environment where the net resource-production-rate is higher. If the AP is following a trajectory $\vb{X}(t)$ such that $\vb{X}(t_0)=\vb{x}_0$, and we assume the resource landscape is stationary over the timescale of the AP's migration ($\partial_t \mathcal{U} = 0$), the active Hamiltonian becomes a conserved quantity along the path ($\text{d}\mathcal{E}/\text{d}t = \partial_t \mathcal{E} = 0$), thus giving us the exact form of Hamiltonian as:

\begin{equation}
    \label{newton}
    \mathcal{E}=\mathcal{U}(\vb{x}_0) = \frac{{\vb{P}}^2(\vb{X}(t))}{2c_d} + \mathcal{U}(\vb{X}(t))
\end{equation}

While the meaning of Eq.~(\ref{newton}) is classically well known, in the context of the AP, it translates to the fact that if a creature migrates from a hostile environment to a fertile one, the change in it's power-expenditure will be used to make it migrate faster in search of more resources as long as the creature does not change its lifestyle, or more appropriately, its base-metabolic-rate (BMR). The last step in the quantization is replacing the classical quantities by their respective operators that act on the wavefunction, and we straightforwardly obtain the Active-Schrödinger 's equation as:

\begin{equation}
    \label{born_SE}
      \hat{\mathcal{E}}\psi =\frac{\hat{{\vb{P}}}^2}{2c_d} \psi + \hat{\mathcal{U}}\psi
\end{equation}

From section [\ref{sub_CPA}], the operators whose eigenvalues are $\hbar_a\omega$ and $\hbar_a\vb{k}$ are $\hat{\mathcal{E}}=i\hbar_a\partial_t$ and $\hat{\vb{P}}=-i\hbar_a\nabla$, respectively and the ``potential energy'' operator, $\hat{\mathcal{U}}=\mathcal{U}(\vb{x},t)$, is simply a scalar multiplier in spatial representations. One must remember that Eq.~(\ref{born_SE}) is valid only for time-scales where the amount of living matter is constant. Cellular multiplication, procreation and death necessarily involve non-Hermitian evolution of the wavefunction, which is not the case with our Active-Schrödinger 's equation. Spatial variations of temperature, CO$_2$/O$_2$ levels in air, nitrogen in the soil, etc., are the factors that influence the BMR of all living organisms and must enter the dynamics through $\mathcal{U}$, which is the power-expenditure necessary for bare survival. Just like traditional quantum-mechanics, solving the wavefunction ab-initio becomes impossible for more than a few interacting APs. However, one can still celebrate the underlying math that remains constant both at the smallest scale and the macro scales for not only inorganic or organic matter, but also the living matter. 

This completes our formulation of AQM, with the caveat that the resource fields, $\phi$, $\Phi$, or $\mathcal{U}$, are externally prescribed onto the system and do not autonomously co-evolve with it. Note that the governing equations, Eqs.~(\ref{q1},\ref{NS2},\ref{NS1}) are linear in $\psi$, and in the low-$Re$ regime, they are also linear in $\vb{v}$ and $p$. The burden of introducing nonlinearity into the system's dynamics is therefore transferred to the evolution of operators such as $\hat{\mathcal{E}}$ and $\hat{\vb{P}}$. Similar to traditional quantum mechanics, this nonlinearity may be further resolved by decomposing $\psi$ into the eigenmodes of relevant operators. The time and translation generators are related to time and position through their respective spatiotemporal derivatives, thus, up to multiplicative constants, the energy-expenditure and momentum-generation-rate representation is the Fourier transform of the spatiotemporal representation of the wavefunction.

Crucially, the equation governing the evolution of $\psi$ is first-order in both time and space, making it inherently suited for relativistic treatment in any Euclidean or non-Euclidean geometry. Notably, this conceptual bridge has already been utilized in the literature concerning analogue black holes. Here, we demonstrate that the evolution of active creatures is similarly analogous to the linear Algaebra of quantum mechanics, with the primary distinction being that the classical definition of action [J$\cdot$s] is replaced by activity [J].

\section{Quantized Active Particles:}\label{sec3}

Having established how continuous energy consumption, fluid advection, and force generation collectively govern the spatiotemporal evolution of living matter, we now anchor this theoretical framework in physical reality. In this section, we highlight bio-physical systems that satisfy these postulates.

We begin with force monopoles, namely APs whose dominant modes of motion are crawling and climbing (see Fig. \ref{nfig_3}a). Recall that $c_d \vb{v}_a= q\hbar_a \nabla\phi$. The AP orients and scales its motion along the direction of the maximum difference between available resources and basal metabolic requirements, i.e., $\hbar_a\nabla\theta$.  Such APs must always be accompanied by an external momentum sink, a requirement captured by the first term in Eq.~(\ref{dist_MP}). Since $q$ and $\hbar_a$ are constants, the entire force-monopole contribution reduces to the gradient of a conservative scalar field, which if not EMF or gravitational field (fields that operate on action minimization), must be the resource field that minimizes the activity.

One prominent exception to this rule, where (pseudo) particles travel as force-monopoles without external force are the topological defects studied by the field of Active Nematics (see Fig. \ref{nfig_3}a[vii]).  For passive crystals, these defects obey Gauss's law of attraction and ultimately push the nematic ordering towards uniformity through pair annihilation. Nematic fields are known to spontaneously create/annihilate as well as sustain topological defects. The flow around such defects is known to preserve vorticity structure \cite{head2024spontaneous} which perfectly mirrors the force-monopoles for +1/2 defects. Similar to the spinor quantum-mechanics, the object describing nematic fields is not a scalar but a multi-component, nematic-order tensor. 

Illustrated in Fig. \ref{nfig_3}b, next in the polar expansion series is the force-dipole term. When a particle pushes forward with a force, proportional to its velocity $\vb{F}=c_d\vb{v}_a$, the object that is pushed backwards is the fluid itself. If the two force vectors are co-axial, the AP can be simply called a ``stresslet''. We are limiting the scope of our paper to pure translation; however, if one defines $\psi$ to represent spinning APs, spinor wavefunction, angular-momentum-generation-rate operator and ``rotlets'', the force-dipole with non-coaxial force pair become appropriate to be defined. Stresslets are the force-free swimmers, such as  rear-actuated ``pushers'' with multiple flagella like E. coli and single-tailed sperm cells, front-actuated ``pullers'' like the biflagellate Chlamydomonas, and surface-driven cilia swimmers like Paramecium. 

This exhausts the type of APs that can move through the fluid. Force-quadrupole and higher-order moments are predominantly associated with ``shakers'' and decay with distance sharply ($\mathcal{O}(1/x^3)$). Interestingly, in the other direction, source monopoles could very well represent moving (creeping) creatures whose motion is captured by diffusioosmosis, a fuller description of which is left for future research.

\section{Discussion and Conclusion}\label{conclusion}

In this work, we have formulated a probabilistic description of active matter by modifying the traditional Madelung equations and the classical idea of the `principle of least action' to represent activity of living matter that inevitably dissipates the otherwise conserved classical quantities such as energy and momentum. The constant increase in entropy and the constant loss of exergy (useful energy) have traditionally obscured the dynamics dictated by energy-expenditure-rate and momentum-generation-rate, which serve as the time-differentiated counterparts to the classical constants of inertial motion. This shifts our descriptive framework from a category of conserved, static states to a category of continuous, active fluxes. 

In future work, we strongly expect mass, angular-momentum, rotational-energy, etc., to map onto their respective time-differentiated category elements as well. Such mapping is imperative if one aims to describe the dynamics of a broader class of active matter. If the active creatures have mechanisms to store energy, or in other words, posses calorie-dense mass, the creatures must expel and consume mass at the exact same rate under steady state, store more mass under abundance and expel more mass under scarcity. To a reasonable but varying degree of validity, a greater stored mass usually means bigger, bulkier bodies that have a larger drag coefficient (or a larger effective mass of AQM) -- for such systems, an investigation into energy-expenditure-rate and effective-mass equivalence seems appropriate. Furthermore, if such energy or mass is constantly expended to change direction (rotate), to extend our framework (which is valid for Stokeslets and stresslets) to rotlets, a rotational-energy-expenditure-rate and a torque-production term must be added to the active Hamiltonian, which we now reasonably expect to behave similar to rotational energy and angular momentum.

While one of the highlights of AQM is the time-differentiated category of observables, the existence of gauge invariance and the first order coupling to the fluid velocity in a Dirac-like equation, Eq.~(\ref{gauge}), is a major structural advantage of this formulation. If the geometry of the spacetime is not linear, such as in thin and curved bio-films, a non-identity metric tensor can easily be added to shift from vectors to co-vectors and hence, define the inner-product for calculating suitable spatial distances. Furthermore, addition of each new resources modifies the dynamics of the wavefunction by locally modifying the complex phase angle, despite which, Eq.~(\ref{gauge}) remains invariant. This local $U(1)$ gauge symmetry entails that no matter what the starting phase of the creatures is, the dynamics will still follow the same principles of achieving maximum resources in minimum time without halting to death by over-expenditure. The corresponding fluid flow, mimicking gauge theoretic formulation of electromagnetism, is naturally obtained by the active force term that is proportional to the creatures' affinity of resources, creatures' dimensions and the distribution of the creatures. 

The first order wavefunction evolution is well suited for changing reference frames between creatures in non-Euclidean geometry. The active Hamiltonian and the corresponding Active-Schrödinger 's equation (Eq.~(\ref{born_SE})) on the other hand, describes the evolution of the population density of mutually non-interacting active creatures which use energy for pure survival (BMR) and movement to fertile environment. In future work, we expect augmentations to the current framework where the wavefunction is not simply a complex scalar but a multi-component tensor. Such tensor would allow for bleeding of energy from pure motion to other facets of life such as reproduction, metamorphosis, or even storing of accumulated energy for future use. Lastly, unlike traditional quantum-mechanics, the need to obtain a complex valued action $S_a$ and hence, a non-conservative wavefunction is pertinent as all known lifeforms grow, die and disintegrate into nothingness over time. We hope the shift from the physical action $S$ [J-s] to biological activity $S_a$ [J] proves useful to the community in furthering our understanding of living matter.

\appendix
\section{ Derivation for Eq.~(\ref{q1})}

  Take the definition of the wave function. (Eq.~(\ref{eq:A1}) is same as Eq.~(\ref{wf_def}) from the main paper):
\begin{equation}
\psi(\mathbf{x}, t) = R(\mathbf{x}, t) \exp(i\theta(\mathbf{x}, t)) \label{eq:A1}\tag{A1}
\end{equation}

  Apply the flux-conserving derivative $d_t^{(\mathbf{v})} = (\partial_t + \nabla \cdot \mathbf{v})$ to both sides of Eq.~(\ref{eq:A1}). Expanding this using the product rule and vector calculus identities yields:
\begin{align*}
d_t^{(\mathbf{v})} \psi &= \partial_t(R e^{i\theta}) + \nabla \cdot (\mathbf{v} R e^{i\theta}) \\
&= (\partial_t R) e^{i\theta} + i R (\partial_t \theta) e^{i\theta} + e^{i\theta} \nabla \cdot (\mathbf{v} R) + \mathbf{v} R \cdot \nabla (e^{i\theta}) \\
&= \left[ \partial_t R + \nabla \cdot (\mathbf{v} R) \right] e^{i\theta} + i R e^{i\theta} (\partial_t \theta + \mathbf{v} \cdot \nabla \theta)
\end{align*}

  Take the condition provided in the main paper, $d_t^{(\mathbf{v})} R = 0$ (which is exactly $\partial_t R + \nabla \cdot (\mathbf{v} R) = 0$), and substitute it into the first bracket of the expanded derivative. This bracket vanishes, leaving:
\begin{equation}
d_t^{(\mathbf{v})} \psi = i \psi (\partial_t \theta + \mathbf{v} \cdot \nabla \theta) \label{eq:A2}\tag{A2}
\end{equation}

  Take the expanded definition of the phase. (Eq.~(\ref{eq:A3}) is same as Eq.~(\ref{phi_split}) from the main paper):
\begin{equation}
\theta(\mathbf{x}, t) = \frac{q}{\hbar_a} \phi(\mathbf{x}, t) + \theta_0(t) \label{eq:A3}\tag{A3}
\end{equation}

  Substitute the phase definition from Eq.~(\ref{eq:A3}) into our derivative expression in Eq.~(\ref{eq:A2}), separating the operators conceptually as presented in the paper. (Eq.~(\ref{eq:A4}) is same as Eq.~(\ref{phi_rate_1}) from the main paper):
\begin{equation}
d_t^{(\mathbf{v})} \psi = i \left( \frac{q}{\hbar_a} d_t^{(\mathbf{v})} \phi + d_t^{(0)} \theta_0 \right) \psi \label{eq:A4}\tag{A4}
\end{equation}

  To get the final physical form, take Eq.~(\ref{eq:A2}) again. Let us clearly define the components of the derivatives of $\theta$ based on Eq.~(\ref{eq:A3}). For the time-dependent phase $\theta_0(t)$, define its partial time derivative as $\partial_t \theta_0(t) = -\omega_0$ and note its spatial gradient is zero ($\nabla \theta_0(t) = 0$). For the $\phi$-related terms, take the phase relations defined in the text (Eqs. \ref{eq:A5} and \ref{eq:A6} are same as Eq.~(\ref{time_gen}) and Eq. (\ref{pos_gen}) from the main paper):
\begin{align}
\partial_t \left(\frac{q}{\hbar_a} \phi(\mathbf{x}, t)\right) &= -\omega(\mathbf{x}, t) \label{eq:A5}\tag{A5} \\
\nabla \left(\frac{q}{\hbar_a} \phi(\mathbf{x}, t)\right) &= \mathbf{k}(\mathbf{x}, t) \label{eq:A6}\tag{A6}
\end{align}

  Take the resulting full time derivative $\partial_t \theta(\mathbf{x}, t) = -\omega(\mathbf{x}, t) - \omega_0$, and substitute it into $\partial_t \theta$ in Eq.~(\ref{eq:A2}).

  Take the full spatial gradient $\nabla \theta(\mathbf{x}, t) = \mathbf{k}(\mathbf{x}, t) + 0 = \mathbf{k}(\mathbf{x}, t)$, and substitute it into $\nabla \theta$ in Eq.~(\ref{eq:A2}).

  Combining these substitutions into Eq.~(\ref{eq:A2}) gives the final result. (Eq.~(\ref{eq:A7}) is same as Eq.~(\ref{q1}) from the main paper):
\begin{equation}
(\partial_t + \nabla \cdot \mathbf{v}) \psi = i (-\omega_0 - \omega + \mathbf{v} \cdot \mathbf{k}) \psi \label{eq:A7}\tag{A7}
\end{equation}

\newpage
\section{Derivation of Eq.~(\ref{point_MP}) from a general distribution}
Take the general continuous force distribution $\mathbf{f}_a=\mathbf{e}_i(f^i)_a(\mathbf{x})$ centered around a reference point $\mathbf{X}$, and take its Fourier transform with respect to the wavenumber $\mathbf{q}$ (not to be confused with the affinity of APs to the resources represented by the non-bold, regular $q$):
\begin{equation}
\tilde{f}^i_a(\mathbf{q}) = \int (f^i)_a(\mathbf{x}) \exp(-i \mathbf{q} \cdot \mathbf{x}) d\mathbf{x} \label{eq:B1}\tag{B1}
\end{equation}

To localize the influence around the center of action $\mathbf{X}$, take the identity $\exp(-i \mathbf{q} \cdot \mathbf{x}) = \exp(-i \mathbf{q} \cdot \mathbf{X}) \exp(-i \mathbf{q} \cdot (\mathbf{x} - \mathbf{X}))$ and substitute it into the integrand of Eq.~(\ref{eq:B1}):
\begin{equation}
\tilde{f}^i_a(\mathbf{q}) = \exp(-i \mathbf{q} \cdot \mathbf{X}) \int (f^i)_a(\mathbf{x}) \exp(-i \mathbf{q} \cdot (\mathbf{x} - \mathbf{X})) d\mathbf{x} \label{eq:B2}\tag{B2}
\end{equation}

Take the Taylor expansion of the exponential term $\exp(-i \mathbf{q} \cdot (\mathbf{x} - \mathbf{X}))$ evaluated for small $|\mathbf{x} - \mathbf{X}|$:
\begin{equation}
\begin{aligned}
&\exp(-i \mathbf{q} \cdot (\mathbf{x} - \mathbf{X})) = 1 - i q_j (x^j - X^j) \\
&\quad + \frac{1}{2} (-i q_j)(-i q_k) (x^j - X^j)(x^k - X^k) - \dots
\end{aligned} \label{eq:B3}\tag{B3}
\end{equation}

Substitute the expanded series from Eq.~(\ref{eq:B3}) back into the integral for the force distribution in Eq.~(\ref{eq:B2}):
\begin{equation}
\begin{aligned}
&\tilde{f}^i_a(\mathbf{q}) = \exp(-i \mathbf{q} \cdot \mathbf{X}) \int (f^i)_a(\mathbf{x}) \Big[ 1 - i q_j (x^j - X^j) \\
&\quad - \frac{1}{2} q_j q_k (x^j - X^j)(x^k - X^k) - \dots \Big] d\mathbf{x}
\end{aligned} \label{eq:B4}\tag{B4}
\end{equation}

Take the definitions of the force multipole moments (the force monopole $F^i$, the force dipole $D^{ij}$, and the force quadrupole $Q^{ijk}$):
\begin{align}
F^i &= \int (f^i)_a(\mathbf{x}) d\mathbf{x} \notag \\
D^{ij} &= \int (x^j - X^j) (f^i)_a(\mathbf{x}) d\mathbf{x} \label{eq:B5}\tag{B5} \\
Q^{ijk} &= \int (x^j - X^j)(x^k - X^k) (f^i)_a(\mathbf{x}) d\mathbf{x} \notag
\end{align}

Substitute the defined moments from Eq.~(\ref{eq:B5}) into the expanded Fourier representation in Eq.~(\ref{eq:B4}):
\begin{equation}
\tilde{f}^i_a(\mathbf{q}) = \exp(-i \mathbf{q} \cdot \mathbf{X}) \left[ F^i - i q_j D^{ij} - \frac{1}{2} q_j q_k Q^{ijk} - \dots \right] \label{eq:B6}\tag{B6}
\end{equation}

Revert back to the spatial representation by taking the inverse Fourier transform. Take the standard Fourier properties where multiplication by an exponential in frequency space ($\exp(-i \mathbf{q} \cdot \mathbf{X})$) translates to a spatial shift resulting in a Dirac delta function $\delta(\mathbf{x} - \mathbf{X})$, and multiplication by $(-i q_j)$ corresponds to the partial spatial derivative $\partial_j$.

Take the definition of the inverse Fourier transform:
\begin{equation}
(f^i)_a(\mathbf{x}) = \frac{1}{(2\pi)^3} \int \tilde{f}^i_a(\mathbf{q}) \exp(i \mathbf{q} \cdot \mathbf{x}) d\mathbf{q} \label{eq:B7}\tag{B7}
\end{equation}
and substitute the expanded representation of $\tilde{f}^i_a(\mathbf{q})$ from Eq.~(\ref{eq:B6}) into Eq.~(\ref{eq:B7}) to prepare for term-by-term integration.

\begin{equation}
\begin{aligned}
&(f^i)_a(\mathbf{x}) = F^i \delta(\mathbf{x} - \mathbf{X}) - D^{ij} \partial_j \delta(\mathbf{x} - \mathbf{X}) \\
&\quad + \frac{1}{2} Q^{ijk} \partial_j \partial_k \delta(\mathbf{x} - \mathbf{X}) - \dots
\end{aligned} \label{eq:B8}\tag{B8}
\end{equation}

Applying these inverse transforms term by term yields the final spatial field. (Eq.~(\ref{eq:B8}) is same as  Eq.~(\ref{point_MP}) from the main paper).
\bibliography{our_bibliography}

@article{solon2015active,
  title={Active brownian particles and run-and-tumble particles: A comparative study},
  author={Solon, Alexandre P and Cates, Michael E and Tailleur, Julien},
  journal={The European Physical Journal Special Topics},
  volume={224},
  number={7},
  pages={1231--1262},
  year={2015},
  publisher={Springer}
}

@article{willems2025run,
  title={Run-and-tumble dynamics of active giant vesicles},
  author={Willems, Vivien and Baron, Alexandre and Fernandez-Matoz, Daniel and Wolfisberg, Gianna and Baret, Jean-Christophe and Dufresne, Eric and Alvarez, Laura},
  journal={Soft Matter},
  volume={21},
  number={31},
  pages={6175--6185},
  year={2025},
  publisher={The Royal Society of Chemistry}
}

@article{quillen2020boids,
  title={Boids in a loop: Self-propelled particles within a flexible boundary},
  author={Quillen, AC and Smucker, JP and Peshkov, A},
  journal={Physical Review E},
  volume={101},
  number={5},
  pages={052618},
  year={2020},
  publisher={APS}
}

@article{baconnier2025self,
  title={Self-aligning polar active matter},
  author={Baconnier, Paul and Dauchot, Olivier and D{\'e}mery, Vincent and D{\"u}ring, Gustavo and Henkes, Silke and Huepe, Cristi{\'a}n and Shee, Amir},
  journal={Reviews of Modern Physics},
  volume={97},
  number={1},
  pages={015007},
  year={2025},
  publisher={APS}
}

@article{ngo2014large,
  title={Large-scale chaos and fluctuations in active nematics},
  author={Ngo, Sandrine and Peshkov, Anton and Aranson, Igor S and Bertin, Eric and Ginelli, Francesco and Chat{\'e}, Hugues},
  journal={Physical review letters},
  volume={113},
  number={3},
  pages={038302},
  year={2014},
  publisher={APS}
}

@article{thampi2016active,
  title={Active turbulence in active nematics},
  author={Thampi, SumeshP and Yeomans, JuliaM},
  journal={The European Physical Journal Special Topics},
  volume={225},
  number={4},
  pages={651--662},
  year={2016},
  publisher={Springer}
}

@article{reddiger2023towards,
  title={Towards a mathematical theory of the Madelung equations: Takabayasi’s quantization condition, quantum quasi-irrotationality, weak formulations, and the Wallstrom phenomenon},
  author={Reddiger, Maik and Poirier, Bill},
  journal={Journal of Physics A: Mathematical and Theoretical},
  volume={56},
  number={19},
  pages={193001},
  year={2023},
  publisher={IOP Publishing}
}

@article{waegell2024toward,
  title={Toward local Madelung mechanics in spacetime},
  author={Waegell, Mordecai},
  journal={Quantum Studies: Mathematics and Foundations},
  volume={11},
  number={3},
  pages={409--426},
  year={2024},
  publisher={Springer}
}

@article{marchetti2013hydrodynamics,
  title={Hydrodynamics of soft active matter},
  author={Marchetti, M Cristina and Joanny, Jean-Fran{\c{c}}ois and Ramaswamy, Sriram and Liverpool, Tanniemola B and Prost, Jacques and Rao, Madan and Simha, R Aditi},
  journal={Reviews of modern physics},
  volume={85},
  number={3},
  pages={1143--1189},
  year={2013},
  publisher={APS}
}

@article{Yeomans08122025,
author = {Julia M. Yeomans and Saraswat Bhattacharyya and Mehrana R. Nejad},
title = {Active nematics: a new approach to mechanobiology?},
journal = {Liquid Crystals},
volume = {52},
number = {14-15},
pages = {1465--1473},
year = {2025},
publisher = {Taylor \& Francis},
doi = {10.1080/02678292.2025.2517180},
}

@article{loewe2018flocking,
  title={Flocking from a quantum analogy: spin--orbit coupling in an active fluid},
  author={Loewe, Benjamin and Souslov, Anton and Goldbart, Paul M},
  journal={New Journal of Physics},
  volume={20},
  number={1},
  pages={013020},
  year={2018},
  publisher={IOP Publishing}
}

@article{te2023microscopic,
  title={From a microscopic inertial active matter model to the Schr{\"o}dinger equation},
  author={Te Vrugt, Michael and Frohoff-H{\"u}lsmann, Tobias and Heifetz, Eyal and Thiele, Uwe and Wittkowski, Raphael},
  journal={Nature Communications},
  volume={14},
  number={1},
  pages={1302},
  year={2023},
  publisher={Nature Publishing Group UK London}
}

@article{houston2024spontaneous,
  title={Spontaneous flows and quantum analogies in heterogeneous active nematic films},
  author={Houston, Alexander JH and Mottram, Nigel J},
  journal={Communications Physics},
  volume={7},
  number={1},
  pages={375},
  year={2024},
  publisher={Nature Publishing Group UK London}
}

@article{ray2014casimir,
  title={Casimir effect in active matter systems},
  author={Ray, D and Reichhardt, C and Reichhardt, CJ Olson},
  journal={Physical Review E},
  volume={90},
  number={1},
  pages={013019},
  year={2014},
  publisher={APS}
}

@article{burgardt2026quantum,
  title={Quantum-enabled active matter at the atomic scale},
  author={Burgardt, Sabrina and Fe{\ss}, Julian and Guthmann, Alexander and Hiebel, Silvia and Mukhopadhyay, Aritra K and Lee, Sangyun and Vrugt, Michael te and Liebchen, Benno and L{\"o}wen, Hartmut and Wittkowski, Raphael and others},
  journal={arXiv preprint arXiv:2606.24615},
  year={2026}
}

@article{antonov2026modeling,
  title={Modeling dissipation in quantum active matter},
  author={Antonov, Alexander P and Lee, Sangyun and Liebchen, Benno and L{\"o}wen, Hartmut and Melles, Jannis and Morigi, Giovanna and Tuchkov, Yehor and te Vrugt, Michael},
  journal={Physical Review A},
  volume={114},
  number={1},
  pages={012201},
  year={2026},
  publisher={APS}
}

@article{reynolds2022stochastic,
  title={Stochastic modelling of bird flocks: accounting for the cohesiveness of collective motion},
  author={Reynolds, Andy M and McIvor, Guillam E and Thornton, Alex and Yang, Patricia and Ouellette, Nicholas T},
  journal={Journal of the Royal Society Interface},
  volume={19},
  number={189},
  pages={20210745},
  year={2022}
}

@article{garrity2010running,
  title={Running hot and cold: behavioral strategies, neural circuits, and the molecular machinery for thermotaxis in C. elegans and Drosophila},
  author={Garrity, Paul A and Goodman, Miriam B and Samuel, Aravinthan D and Sengupta, Piali},
  journal={Genes \& development},
  volume={24},
  number={21},
  pages={2365--2382},
  year={2010},
  publisher={Cold Spring Harbor Laboratory Press}
}

@article{wadhams2004making,
  title={Making sense of it all: bacterial chemotaxis},
  author={Wadhams, George H and Armitage, Judith P},
  journal={Nature reviews Molecular cell biology},
  volume={5},
  number={12},
  pages={1024--1037},
  year={2004},
  publisher={Nature Publishing Group UK London}
}

@article{wilde2017light,
  title={Light-controlled motility in prokaryotes and the problem of directional light perception},
  author={Wilde, Annegret and Mullineaux, Conrad W},
  journal={FEMS microbiology reviews},
  volume={41},
  number={6},
  pages={900--922},
  year={2017},
  publisher={Oxford University Press}
}

@article{nakane2023rheotaxis,
  title={Rheotaxis in Mycoplasma gliding},
  author={Nakane, Daisuke},
  journal={Microbiology and Immunology},
  volume={67},
  number={9},
  pages={389--395},
  year={2023},
  publisher={Wiley Online Library}
}

@inbook{Lauga_2020, place={Cambridge}, series={Cambridge Texts in Applied Mathematics}, title={Hydrodynamics of Slender Filaments}, booktitle={The Fluid Dynamics of Cell Motility}, publisher={Cambridge University Press}, author={Lauga, Eric}, year={2020}, pages={77–96}, collection={Cambridge Texts in Applied Mathematics}}

@article{ganguly2025hydrodynamics,
  title={Hydrodynamics of squirmers: A review on Stokesian principles, analytical frameworks, and recent advances},
  author={Ganguly, Sourav and others},
  journal={Physics of Fluids},
  volume={37},
  number={12},
  year={2025},
  publisher={AIP Publishing}
}

@article{zottl2023modeling,
  title={Modeling active colloids: From active Brownian particles to hydrodynamic and chemical fields},
  author={Z{\"o}ttl, Andreas and Stark, Holger},
  journal={Annual Review of Condensed Matter Physics},
  volume={14},
  number={1},
  pages={109--127},
  year={2023},
  publisher={Annual Reviews}
}

@book{griffiths2008introduction,
  author    = {Griffiths, David},
  title     = {Introduction to Elementary Particles},
  edition   = {2nd},
  year      = {2008},
  publisher = {Wiley-VCH},
  address   = {Weinheim},
  chapter   = {10}
}

@article{head2024spontaneous,
  title={Spontaneous self-constraint in active nematic flows},
  author={Head, Louise C and Dore, Claire and Keogh, Ryan R and Bonn, Lasse and Negro, Giuseppe and Marenduzzo, Davide and Doostmohammadi, Amin and Thijssen, Kristian and L{\'o}pez-Le{\'o}n, Teresa and Shendruk, Tyler N},
  journal={Nature Physics},
  volume={20},
  number={3},
  pages={492--500},
  year={2024},
  publisher={Nature Publishing Group UK London}
}

@article{lara2026forces,
  title={Forces and symmetry breaking of a living meso-swimmer},
  author={Lara, Rafael A and Sharadhi, N and Huttunen, Anna AL and Ansas, Lotta and Rislakki, Ensio JG and Bessa, Guilherme M and Backholm, Matilda},
  journal={Communications Physics},
  volume={9},
  number={1},
  pages={53},
  year={2026},
  publisher={Nature Publishing Group UK London}
}

@article{walker2020computer,
  title={Computer-assisted beat-pattern analysis and the flagellar waveforms of bovine spermatozoa},
  author={Walker, Benjamin J and Phuyal, Shiva and Ishimoto, Kenta and Tung, Chih-Kuan and Gaffney, Eamonn A},
  journal={Royal Society open science},
  volume={7},
  number={6},
  pages={200769},
  year={2020}
}

@article{moreau2024minimal,
  title={Minimal design of a synthetic cilium},
  author={Moreau, Cl{\'e}ment and Walker, Benjamin J and Poon, Rebecca N and Soto, Daniel and Goldman, Daniel I and Gaffney, Eamonn A and Wan, Kirsty Y},
  journal={Physical Review Research},
  volume={6},
  number={4},
  pages={L042061},
  year={2024},
  publisher={APS}
}

@article{wan2014rhythmicity,
  title={Rhythmicity, recurrence, and recovery of flagellar beating},
  author={Wan, Kirsty Y and Goldstein, Raymond E},
  journal={Physical review letters},
  volume={113},
  number={23},
  pages={238103},
  year={2014},
  publisher={APS}
}

@article{lighthill1952sound,
  title={On sound generated aerodynamically. I. General theory},
  author={Lighthill, Michael James},
  journal={Proceedings of the Royal Society of London. Series A, Mathematical and Physical Sciences},
  pages={564--587},
  year={1952},
  publisher={JSTOR}
}

@article{lighthill1954sound,
  title={On sound generated aerodynamically. II. Turbulence as a source of sound},
  author={Lighthill, Michael James},
  journal={Proceedings of the Royal Society of London. Series A, Mathematical and Physical Sciences},
  pages={1--32},
  year={1954},
  publisher={JSTOR}
}
\end{document}